\documentclass[10pt,a4]{article}

\usepackage{latexsym,amsmath,amscd,amssymb,graphicx}

\begin{document}
\begin{titlepage}

\title{A new view on relativity:\\Part 2. Relativistic dynamics}

\author{Yaakov Friedman\\Jerusalem College of
Technology\\ P.O.B. 16031 Jerusalem 91160\\
Israel\\e-mail:friedman@jct.ac.il}

\date{}
\maketitle

\begin{abstract}

 The Lorentz transformations are represented on the ball of
relativistically admissible velocities by Einstein velocity
addition and rotations. This representation is by projective maps.
The relativistic dynamic equation can be derived by introducing a
new principle which is analogous to the Einstein's Equivalence
Principle, but can be applied for any force. By this principle,
the relativistic dynamic equation is defined by an element of the
Lie algebra of the above representation.

 If we introduce a new dynamic variable, called
symmetric velocity, the above representation becomes a
representation by conformal, instead of projective maps. In this
variable, the relativistic dynamic equation for systems with an
invariant plane, becomes a non-linear analytic equation in one
complex variable. We obtain explicit solutions for the motion of a
charge in uniform, mutually perpendicular electric and magnetic
fields.

By the above principle, we show that the relativistic dynamic
equation for the four-velocity leads to an analog of the
electromagnetic tensor. This indicates that force in special
relativity is described by a differential two-form.

 \textit{PACS}: 03.30.+p ; 03.50-z.

\end{abstract}

\end{titlepage}

\section{Introduction}

In Part 1 we have shown that from the Principle of Relativity
alone, we can infer that there are only two possibilities for
space time transformations between inertial systems: the Galilean
transformations or the Lorentz transformations. In Special
Relativity we use the Lorentz transformation and obtain interval
conservation. We also show that the set of all relativistically
allowed velocities is a ball $D_v \in R^3$ of radius $c$-the speed
of light. We have shown that  similar results hold for
proper-velocity-time transformations between accelerated systems.

If an object moves in an inertial system $K'$ with uniform
velocity $\mathbf{u}$ and $K'$ moves parallel to $K$ with relative
velocity $\mathbf{b}$, then in system $K$ the object has uniform
velocity $\mathbf{b} \oplus \mathbf{u}$, the relativistic sum of
$\mathbf{b}$ and $\mathbf{u},$ defined as
\begin{equation}\label{veladd}
  \mathbf{b} \oplus \mathbf{u}=\frac{ \mathbf{b} +  \mathbf{u}_{\|}+\alpha \mathbf{u}_\bot}{1 +
   \frac{\langle\mathbf{b}|\mathbf{u}\rangle}{c^2}},
\end{equation}
where $\mathbf{u}_{\|}$ denotes
          the component of $\mathbf{u}$ parallel to $\mathbf{b}$,
          $\mathbf{u}_\bot$ denotes
          the component of $\mathbf{u}$ perpendicular  to $\mathbf{b}$
         and $\alpha=\sqrt{1-|\mathbf{b}|^2/c^2}$. This is the well-known
Einstein velocity addition formula. Note that the velocity
addition is \textit{commutative only} for parallel velocities. The
Lorentz transformation  preserves the velocity ball $D_v$ and acts
on it by
\begin{equation}\label{boostvelball}
 \varphi _{\mathbf{b}}(\mathbf{u})=\mathbf{b} \oplus \mathbf{u}.
\end{equation}
 It can be shown \cite{F04}
that the map $\varphi _{\mathbf{b}}$ is a projective (preserving
line segments) automorphism of $D_v$.

We denote by $Aut _p(D_v)$ the group of all projective
automorphisms of the domain $D_v$. The map $\varphi _{\mathbf{b}}$
belongs to $Aut _p(D_v)$.  Let $\psi$ be any projective
automorphism of $D_v$. Set $\mathbf{b}=\psi (0)$ and $U=
\varphi_{\mathbf{b}}^{-1}\psi$. Then $U$ is an isometry and
represented by an orthogonal matrix. Thus, the group $Aut _p(D_v)$
of all projective automorphisms has the characterization
\begin{equation}\label{AutD0}
 Aut _p(D_v)=\{\varphi_{\mathbf{b}}U :
  \mathbf{b} \in D_v, \; U\in O(3)\}.
\end{equation}
This group represents the velocity transformation between two
arbitrary inertial systems and provides a representation of the
Lorentz group.

 Now we are going to adapt Newton's classical dynamics law
  \[\mathbf{F}=m\mathbf{a}=m\frac{d\mathbf{v}}{dt}\] to
 special relativity. By definition, a force
 generates a change of velocity. Since in special relativity the velocity
 must remain in $D_v$, the changes caused by the force cannot take the
 velocity out of $D_v$. This implies that on the boundary of the domain
 $D_v$, the changes caused by the force cannot have a non-zero
 component normal to the boundary of the domain and facing outward.
  One of the common ways to solve this problem is
 to assume that the mass $m$ approaches infinity as the velocity
 approaches the boundary of $D_v$.

 We consider the mass of an object to be an intrinsic
 characteristic of the object. We therefore keep the mass constant and
 equal to the so-called rest-mass $m_0$. Under such an assumption we
 must give up the property that the force on an object is
 independent of the velocity of the object, since such a force
 would take the velocity out of $D_v$. Note that also in
 non-relativistic mechanics we have forces which depend on the
 velocity, like friction and the magnetic force.

 To derive the relativistic dynamics equation, we must introduce a
 new axiom which  will allow us to derive such an equation. For
 alternative axioms used by others, see Rindler
  \cite{rindler}, p.109. Based on this new axiom, we will derive a
  relativistic dynamics equation. Our equation  agrees with the known relativistic dynamics equation obtained by different assumptions. The difference will be only in the interpretation and the derivation of the equation.

\section{Extended Partial Equivalence Principle \textemdash $EP^2$}

We base our additional axiom for relativistic dynamics on
Einstein's Equivalence Principle. In the context of flat
space-time, the \textit{Principle of Equivalence} states that
``the laws of physics have the same form in a uniformly
accelerated system as they have in an unaccelerated inertial
system in a uniform gravitational field." This means that the
evolution  of an object in an inertial system $K$ under a uniform
gravitational field or gravitational force is the same as the free
motion of the object in the system $K'$ moving with uniform
acceleration with respect to $K$.

 We denote the relative velocity
of the system $K'$ with respect to $K$ caused by this uniform
acceleration by $\mathbf{b}(t)$ and assume that $\mathbf{b}(0)=0$.
 Since in the system $K'$ the motion of the object is free, its velocity
$\mathbf{u}(t')$ there is constant and is equal to its initial
velocity $\mathbf{u}(t')=\mathbf{v}_0$. By (\ref{boostvelball}),
the velocity of the object in system $K$ is
\begin{equation}\label{forceonvelball}
 \mathbf{v}(t)=\mathbf{b}(t)\oplus
\mathbf{u}(t')=\mathbf{b}(t)\oplus
\mathbf{v}_0=\varphi_{\mathbf{b}(t)}(\mathbf{v}_0).
\end{equation}
In particular, $\mathbf{b}(t)$ is the velocity at time $t$ of an
object moving  under the force of our gravitational field which
was at rest at $t=0$.

From this observation, we see that the Principle of Equivalence
provides a connection between the action of a force on an object
with zero initial condition and its action on an object with
nonzero initial condition. Moreover, equation
(\ref{forceonvelball}) implies that a uniform gravitational force
in Special Relativity defines an evolution on the velocity ball
$D_v$ which is given by a differentiable curve $g
(t)=\varphi_{\mathbf{b}(t)}\in Aut _p(D_v),$  with $g
(0)=\varphi_{0,I}$-the identity of $Aut _p(D_v)$. Thus, from the
definition of the Lie algebra $aut_p(D_v)$ as generators of such
curves, we conclude that the action of a uniform gravitational
field on the velocity ball $D_v$ is given by an element of the Lie
algebra $aut_p(D_v)$.

We \textit{extend} the Equivalence Principle to a form which will
make it valid for \textit{any} force, not only  gravity  and call
this the ``Extended Partial Equivalence Principle" - $EP^2$ for
short. The statement of this principle is: \textit{The evolution
of an object in an inertial system under a uniform force is the
same as a free evolution of the same (or similar) object in a
uniformly accelerated system.} Since the action of the
gravitational force on an object is independent of the object's
properties, the $EP^2$ for the gravitational force holds  for
\textit{any} object, not only for the same one.

According to the above argument, formula (\ref{forceonvelball})
will hold for any force satisfying $EP^2$, not only the gravity
satisfying $EP$. This means that the velocity of an object under a
uniform force satisfying $EP^2$ in relativistic dynamics is
\begin{equation}\label{initialcondgenforeEP}
\mathbf{v}(t)=\mathbf{b}(t)\oplus \mathbf{v}_0,
\end{equation}
where $\mathbf{b}(t)$ is the velocity evolution of a similar
object with zero initial velocity and $\mathbf{v}_0$ is the
initial velocity of the object. It can be shown that the solution
of the usual relativistic dynamics equation satisfies this
property. Also, as above, the action of any uniform force (on
given objects) on the velocity ball $D_v$ is given by an element
of the Lie algebra $aut_p(D_v)$.

Note that forces satisfying the $EP^2$ do not generate rotations
and thus are represented by a subset of $aut_p(D_v)$ which is not
a Lie algebra. Thus, in order to obtain a Lorentz invariant
relativistic dynamic equation we must assume that a force can be
represented by an arbitrary element of the Lie algebra
$aut_p(D_v)$. This will allow the force to have a rotational
component as well. In the next section, we derive the Relativistic
Dynamic equation for forces satisfying $EP^2$ implying that they
are elements of $aut_p(D_v)$.

\section{Relativistic Dynamics on the velocity ball}

 To define the elements of $aut_p(D_v)$, consider
differentiable curves $g (t)$ from a neighborhood $I_0$ of $0$
into $Aut _p(D_v)$, with $g (0)=\varphi_{0,I}$, the identity of
$Aut _p(D_v)$. According to (\ref{AutD0}), any such $g(t)$ has the
form
\begin{equation}\label{gamas}
g(t)=\varphi_{\mathbf{b}(t)}U(t),
\end{equation}
where $\mathbf{b}:I_0 \rightarrow D_v$ is a differentiable
function satisfying $\mathbf{b}(0)=\mathbf{0}$ and
$U(t):I_0\rightarrow O(3)$ is differentiable and satisfies
$U(0)=I$. We denote by $\delta$ the element of $aut_p(D_v)$
generated by $g(t)$. By direct calculation (see \cite{F04}, p.35),
we get
\begin{equation}\label{autD}
\delta (\mathbf{v})=\frac {d}{dt}g (t)(\mathbf{v})\Bigr|_{t=0}=
\mathbf{E}+A\mathbf{v}-c^{-2}\langle\mathbf{v}|\mathbf{E}\rangle\mathbf{v},
\end{equation}
where $\mathbf{E}=\mathbf{b}'(0)\in R^3$ and $A=U'(0)$ is a
$3\times 3$ skew-symmetric matrix $\begin{pmatrix}
  0 & a_{12} & a_{13} \\
  -a_{12} & 0 & a_{23} \\
  -a_{13} & -a_{23} & 0
\end{pmatrix}$. Defining
$\frac{\mathbf{B}}{c}=\begin{pmatrix}
  a_{23} \\
  -a_{13} \\
  a_{12}
\end{pmatrix}$, we have
\begin{equation}\label{AandB}
  A\mathbf{v}=\mathbf{v}\times \frac{\mathbf{B}}{c}=\frac{\mathbf{v}}{c}\times \mathbf{B},
\end{equation}
where $\times$ denotes the vector product in $R^3$. Thus, the
\index{Lie algebra!$aut_p(D_v)$} Lie algebra
\begin{equation}\label{autD1}
 aut_p(D_v)=\{\delta _{\mathbf{E},\mathbf{B}} :\mathbf{E},\mathbf{B} \in R^3  \},
\end{equation}
where $\delta _{\mathbf{E},\mathbf{B}}:D_v \rightarrow R^3$ is the
vector field defined by
\begin{equation}\label{deltaEB}
 \delta _{\mathbf{E},\mathbf{B}}(\mathbf{v})=
\mathbf{E}+\frac{\mathbf{v}}{c}\times
\mathbf{B}-c^{-2}\langle\mathbf{v}\,|\mathbf{E}\rangle\mathbf{v}.
\end{equation}

Note that any $\delta_{\mathbf{E},\mathbf{B}} (\mathbf{v})$ is a
polynomial in $\mathbf{v}$ of degree less than or equal to 2. This
is a general property of the Lie algebra of the automorphism group
of a Bounded Symmetric Domain, see \cite{F04}. The ball $D_v$ is a
Bounded Symmetric Domain with the automorphism group $Aut _p(D_v)$
of projective maps and $aut _p(D_v)$ is its Lie algebra. It is
known that the elements of the Lie algebra of a Bounded Symmetric
Domain are uniquely described by a triple product, called the
$JB^*$ triple product. The elements of $aut _p(D_v)$ transform
between two inertial systems in the same way as the transformation
of the electromagnetic field strength.

Under our assumption, the force is an element of $aut _p(D_v).$
Thus, the equation of evolution of a charged particle with charge
$q$ and \textit{rest-mass} $m_0$  using the generator
$\delta_{\mathbf{E},\mathbf{B}}\in aut _p(D_v)$ is defined by
\begin{equation} m_0\frac{d\mathbf{v}(\tau)}{d\tau
}=q \delta
_{\mathbf{E},\mathbf{B}}(\mathbf{v}(\tau)),\end{equation}  or
\begin{equation}\label{relevolution}
m_0\frac{d\mathbf{v}(\tau)}{d\tau
}=q(\mathbf{E}+\frac{\mathbf{v}(\tau)}{c} \times
\mathbf{B}-c^{-2}\langle\mathbf{v}(\tau)|\mathbf{E}\rangle
\mathbf{v}(\tau)),
\end{equation}
 where $\tau$ is the proper time of the particle. Note that the
 last (quadratic) term in (\ref{relevolution}) keeps the velocity
 inside the ball and we do not need to introduce varying mass.
 It can be shown \cite{F04} that this formula coincides with the
well-known formula
$$\frac{d(m\mathbf{v})}{dt }=q(\mathbf{E}+ \frac{\mathbf{v}}{c}\times \mathbf{B}).$$

Thus, the flow generated by an electromagnetic field is defined by
elements of the Lie algebra $aut_p(D_v)$, which are, in turn,
vector field polynomials in $\mathbf{v}$ of degree 2. The linear
term of this field comes from the magnetic force, while the
constant and the quadratic terms come from the electric field. If
the electromagnetic field $\mathbf{E},\mathbf{B}$ is constant,
then for any given $\tau$, the solution of (\ref{relevolution}) is
an element $\varphi _{\mathbf{b}(\tau ),U(\tau)}\in Aut _p (D_v)$
and the set of such elements form a one-parameter subgroup of $Aut
_p (D_v)$. This subgroup is a geodesic of the metric invariant
under the group.

If we set $\mathbf{B}=0$ and denote $\mathbf{F}=q\mathbf{E},$ we
obtain the dynamics equation of evolution in relativistic
\textit{mechanics}. Thus, also in relativistic mechanics the force
is defined by an element of  $aut_p(D_v)$.

If the electromagnetic field is not uniform, it is defined by
$\mathbf{E}(t,\mathbf{r})$ and $\mathbf{B}(t,\mathbf{r})$ which
are dependent on space and time. In this case, the action of the
field is on a fibre-bundle with  Minkowski space-time as base and
$D_v$ as the fibre. The field acts on the fibre over the point
$(t,\mathbf{r})$ as $\delta
_{\mathbf{E}(t,\mathbf{r}),\mathbf{B}(t,\mathbf{r})}(\mathbf{v}),$
defined by (\ref{deltaEB}).

\section{Symmetric velocity dynamics}

Explicit solution of the evolution equation (\ref{relevolution})
exists only for constant electric $\mathbf{E}$ or constant magnetic
$\mathbf{B}$ fields. If both fields are present, even in the case where
there is an invariant plane and the problem can be reduced to
one complex variable, there are no direct explicit solutions. The
reason for this is that equation  (\ref{relevolution}) is not
complex analytic. Complex analyticity is connected with conformal
maps, while the transformations on the velocity ball are
projective. All currently known explicit solutions
\cite{Baylis},\cite{Takeuchi02} and \cite{FS} use some
substitutions such that in the new variable the transformations become
conformal.

\begin{figure}[h!]
  \centering
 \scalebox{0.45}{\includegraphics{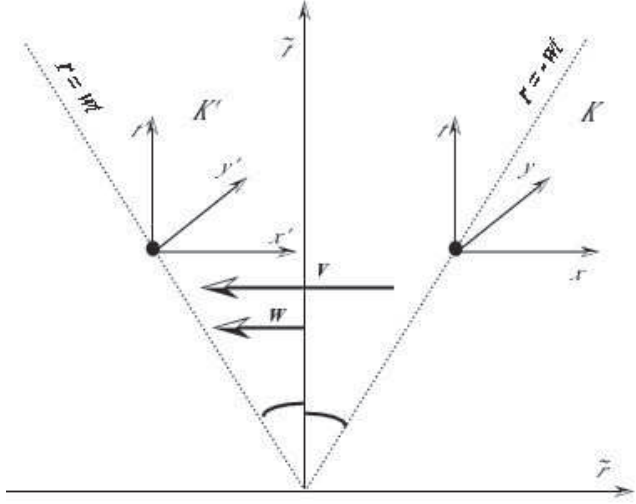}}
  \caption[The physical meaning of symmetric velocity.]
  { The physical meaning of symmetric velocity. Two inertial
  systems $K$ and $K'$ with relative velocity $\mathbf{v}$
between them are viewed from the system connected to their center.
In this system, $K$ and $K'$ are each moving with velocity $\pm
\mathbf{w}$. }\label{svelmeaning2}
\end{figure}

To obtain explicit solutions for motion of a charge in constant,
uniform, and mutually perpendicular electric and magnetic fields,
we associate with any velocity $\mathbf{v}$ a new dynamic variable
called the \textit{symmetric velocity} $\mathbf{w_s}$. The
symmetric velocity $\mathbf{w_s}$ and its corresponding velocity
$\mathbf{v}$ are related by
\begin{equation}\label{velandsymvel}
 \mathbf{v}=\mathbf{w_s}\oplus\mathbf{w_s}=\frac{2\mathbf{w_s}}{1+|\mathbf{w_s}|^2/c^2}.
\end{equation}
The physical meaning of this velocity is explained in Figure
\ref{svelmeaning2}.

Instead of $\mathbf{w_s}$, we shall find it more convenient to use
the unit-free vector $\mathbf{w}=\mathbf{w_s}/c$, which we call
the \textit{s-velocity}. The relation of a velocity $\mathbf{v}$
to its corresponding s-velocity is
\begin{equation}\label{velandsvel}
\mathbf{v}=\Phi(\mathbf{w})=\frac{2c\mathbf{w}}{1+|\mathbf{w}|^2},
\end{equation}
where $\Phi$ denotes the function mapping the s-velocity
$\mathbf{w}$ to its corresponding velocity $\mathbf{v}$. The
s-velocity has some interesting and useful mathematical
properties. The set of all three-dimensional relativistically
admissible s-velocities forms a unit ball
\begin{equation}
D_s = \{\mathbf{w}\in R^3:\; |\mathbf{w}|<1\}.
\end{equation}

Corresponding to the Einstein velocity addition equation, we may
define an addition of s-velocities in $D_s$ such that
\begin{equation}\label{einsteinaddsymaddition}
 \Phi ( \mathbf{b} \oplus _s \mathbf{w})= \Phi(\mathbf{b}) \oplus
  _E  \Phi (\mathbf{w}).
\end{equation}
A straightforward calculation leads to the corresponding equation
for s-velocity addition:
\begin{equation}\label{symveladdition}
  \mathbf{b} \oplus _s \mathbf{w}=
  \frac{(    1+|\mathbf{w}|^2  +2<\mathbf{b}\mid \mathbf{w}> )\mathbf{b}+
             (1-|\mathbf{b}|^2 )\mathbf{w}
        }
        {    1+|\mathbf{b}|^2|\mathbf{w}|^2 +
             2<\mathbf{b}  \mid \mathbf{w}>
        }.
\end{equation}

Equation (\ref{symveladdition}) can be put into a more convenient
form if, for any $\mathbf{b}\in D_s$, we define a map
$\Psi_{\mathbf{b}}:D_s\to D_s$ by
\begin{equation}\label{psia}
 \psi_{\mathbf{b}}(\mathbf{w})\equiv\mathbf{b} \oplus _s
 \mathbf{w}.
\end{equation}
This map is an extension to $D_s \in R^n$ of the M\"{o}bius
addition on the complex unit disc. It defines a \textit{conformal}
map on $D_s$. The motion of a charge in $\mathbf{E} \times
\mathbf{B}$ fields is two-dimensional if the charge starts in the
plane perpendicular to $\mathbf{B}$, and in this case
Eq.(\ref{symveladdition}) for s-velocity addition is somewhat
simpler.  By introducing a complex structure on the plane $\Pi$,
which is perpendicular to $\mathbf{B}$, the disk
$\Delta=D_s\cap\Pi$ can be identified as a unit disc $|w|<1$
called the Poincar\'{e} disc. In this case the s-velocity addition
defined by Eq.(\ref{symveladdition}) becomes
\begin{equation}\label{mobius}
 a \oplus _s w=\psi_{{a}}({w})=\frac{a+w}{1+\overline{a}w},
\end{equation}
which is the well-known M\"{o}bius transformation of the unit
disk.

By using the $s$ velocity we can rewrite ( as in \cite{F04}) the
relativistic Lorentz force equation
$$\frac{d}{dt}(\gamma m\mathbf{v})=q({\bf E} +\mathbf{v}\times
{\bf  B})$$ as
\begin{equation}\label{EB18}
\frac{m_0 c}{q}\,\frac{d\mathbf{w}}{d\tau}
=\left(\frac{1+|\mathbf{w}|^2}{2}\right){\bf{E}}+c\mathbf{w}\times
{\bf{B}}-\mathbf{w}<\mathbf{w}|{\bf{E}}>,
\end{equation}
which is the relativistic Lorentz force equation for the
s-velocity $\mathbf{w}$ as a function of the proper time $\tau$.

We now use Eq.(\ref{EB18}) to find the s-velocity of a charge $q$
in uniform, constant, and mutually perpendicular electric and
magnetic fields.  Since all of the terms on the right hand side of
Eq. (\ref{EB18}) are in the plane perpendicular to ${\mathbf{B}}$,
if $\mathbf{w}$ is in the plane $\Pi$ perpendicular to ${\mathbf{B}}$, then $d{\mathbf{w}}/{d\tau}$ is also in $\Pi$. Consequently, if the
initial s-velocity is in the plane perpendicular to
${\mathbf{B}}$, ${\mathbf{w}}$ will remain in the this plane and
the motion will be two dimensional.

Working in Cartesian coordinates, we choose
\begin{equation}\label{defEB}
\mathbf{E}=(0,E,0),\; \mathbf{B}=(0,0,B),\; \mbox{and}\;
\mathbf{w}=(w_1,w_2,0).
\end{equation}
By introducing a complex structure in $\Pi$ by denoting  $w=w_1 +
iw_2$ the evolution equation Eq.(\ref{EB18}) get the following
simple form:
\begin{equation}\label{EBmain} \frac{dw}{d\tau}
=i\Omega\left(w^2-2\widetilde{B}w+1\right),
\end{equation}
where
\begin{equation}\label{OmegaBtilde}
\Omega\equiv\frac{qE}{2m_0
c}\;\;\mbox{and}\;\;\widetilde{B}\equiv\frac{cB}{E}.
\end{equation}
The solution of Eq.(\ref{EBmain}) is unique for a given initial
condition
\begin{equation}\label{initialconditionEB}
 w(0)=w_0,
\end{equation}
where the complex number $w_0$ represents the initial s-velocity
$\mathbf{w}_0=\Phi^{-1}(\mathbf{v}_0)$ of the charge.

Integrating Eq.(\ref{EBmain}) produces the equation
\begin{equation}\label{intEB2}
\int\frac{dw}{w^2-2\widetilde{B}w+1} =i\Omega \tau +C,
\end{equation}
where the constant $C$ is determined from  the initial condition
(\ref{initialconditionEB}). The way we evaluate this integral
depends upon the sign of the discriminant $4\widetilde{B}^2-4$
associated with the denominator of the integrand. If we define
\begin{equation}\label{discriminantEB}
\Delta\equiv \widetilde{B}^2-1=\frac{(cB)^2-E^2}{E^2},
\end{equation}
then the three cases $E<cB, E=cB\; \mbox{and}\; E>cB$ correspond
to the cases $\Delta$ greater than zero, equal to zero, and less
than zero.
\medskip

\begin{figure}[h!]
  \centering
 \scalebox{0.4}{\includegraphics{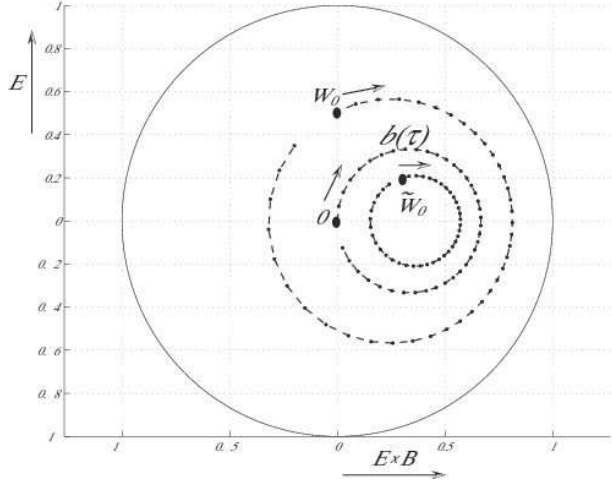}}
  \caption{The trajectories of the s-velocity $w(\tau)$ of a charged
  particle with $q/m=10^7$C/kg
  in a constant uniform fields $E=1$V/m and $cB=1.5$V/m.
  The initial conditions are $w_0=-0.02 +i0.5$ for the first trajectory
  and $\widetilde{w}_0=0.3+i0.2$ for the second. We also draw $b(\tau)$ corresponding to
  $w_0=0$. Note that all the trajectories are circles.}\label{wtaucase3}
\end{figure}

\noindent {\bf Case 1} Consider first the case \begin{equation}
\Delta= ((cB)^2-{E}^2)/{E}^2)>0\Longleftrightarrow
{E}<cB\;\mbox{and}\;\widetilde{B}>1. \end{equation} The
denominator of the integrand in (\ref{intEB2}) can be rewritten as
\begin{equation} w^2-2\widetilde{B}w+1=(w-\alpha _1)(w-\alpha _2),
\end{equation} where $\alpha _1$ and $\alpha _2$ are the real,
positive roots
\begin{equation}\label{EBroots}
\alpha _1=\widetilde{B}-\sqrt{\widetilde{B}^2-1}\;\;\mbox{and}\;\;
\alpha _2=\widetilde{B}+\sqrt{\widetilde{B}^2-1}.
\end{equation}
and the solution then becomes:
\begin{equation}\label{solevoleqncase3}
w(\tau)=\frac{\alpha _1+Ce^{-i\nu\tau}} {1+\alpha
_1Ce^{-i\nu\tau}}=\alpha _1\oplus _s Ce^{-i\nu\tau},
\end{equation}
with \begin{equation}\label{nu}
\nu=\left(\frac{q}{mc}\right)\sqrt{E^2-(cB)^2}.
\end{equation} This equation shows
that in a system K' moving with s-velocity $\alpha _1$ relative to
the lab, the s-velocity of the charge corresponds to circular
motion with initial s-velocity
\begin{equation}\label{initialC}
 C=\psi _{-\alpha _1}(w_0).
\end{equation}
The s-velocity observed in K is shown in Figure \ref{wtaucase3}.

From Eqs.(\ref{velandsvel}) and (\ref{EBroots}) it follows that
the velocity corresponding to s-velocity $\alpha _1$ is
\begin{equation}\label{driftveldef}
  \frac{2c\alpha _1}{1+|\alpha _1|^2}=(E/B)\mathbf{i}=\mathbf{v}_d=v_d\mathbf{i},
\end{equation}
which is the well-known $\mathbf{E}\times \mathbf{B}$ drift
velocity. Applying the map $\Phi$ defined in
Eq.(\ref{einsteinaddsymaddition}) to both sides of
(\ref{solevoleqncase3}), we get
\begin{equation}\label{solevoleqncase3a}
\mathbf{v}(\tau)=\mathbf{v}_d\oplus _E e^{-i\nu\tau}\Phi (C).
\end{equation}
Eq.(\ref{solevoleqncase3a}) says that the total velocity of the
charge,  as a function of the proper time, is the sum of a
constant drift velocity $\mathbf{v}_d= (E/B) \mathbf{i}$ and
circular motion, as expected.

\begin{figure}[h!]
  \centering
 \scalebox{0.4}{\includegraphics{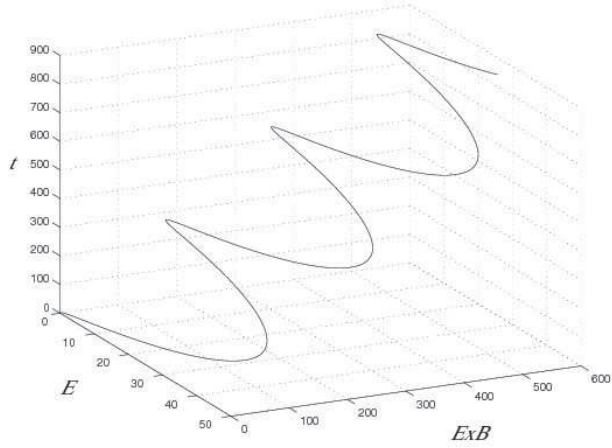}}
  \caption{The world line $\mathbf{r}(\tau),t(\tau)$  of the test particle of Figure \ref{wtaucase3}
   in the same electromagnetic field. The initial velocity
   $\mathbf{v}_0=(0,0,0)$. }\label{worldrase3}
\end{figure}

If we let
\begin{equation}\label{phi(C)decomp}
 \Phi (C)=|\Phi (C)|e^{i\tau_0}=\widetilde{v}_0e^{i\tau_0},
\end{equation}
then the velocity of the charge is
\begin{equation}\label{v(tau)B}
\mathbf{v}(\tau)=\mathbf{v} _d\oplus _E
\widetilde{v}_0e^{-i\nu(\tau-\tau_0)}.
\end{equation}
The position of the charge as a function of the proper time is
\begin{equation}\label{posproptime}
\mathbf{r}(\tau)=\int_0^{\tau}\gamma\mathbf{v}(\tau')\,d\tau' =
\frac{\gamma_dv_d}{\nu}\left(\gamma_d
(\nu\tau-\sin{\nu\tau}),(\cos\nu\tau-1)\right)\end{equation} and
the lab time $t$  as a function of the proper time is
\begin{equation}\label{tastaucase3}
 t(\tau)=\int_0^{\tau}\gamma(\tau')d
 \tau'=\frac{\gamma_d^2}{\nu}\left(\nu\tau-\frac{v_d^2}{c^2}\sin{\nu\tau}\right),
\end{equation}
where $\gamma_d=\gamma(\mathbf{v}_d)$. The world line
$\mathbf{r}(\tau),t(\tau)$ of such test particle is presented on
Figure \ref{worldrase3}.

\medskip

\begin{figure}[h!]
  \centering
 \scalebox{0.35}{\includegraphics{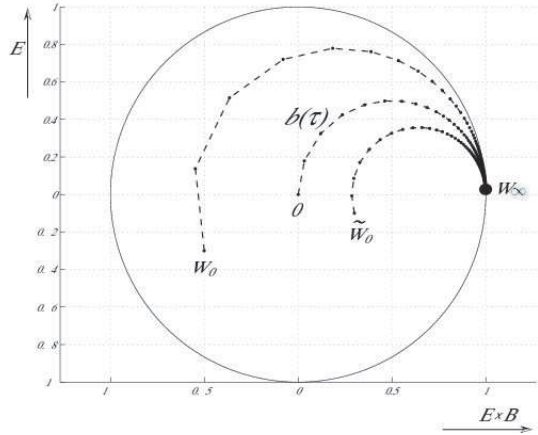}}
  \caption[Charged particle trajectories of the s-velocity $w(\tau)$ for $|{\bf  E}|=c|{\bf  B}|$.]
  {The trajectories of the s-velocity $w(\tau)$ of a charged
  particle with $q/m=10^7$C/kg
  in constant, uniform fields $E=1$V/m and $cB=1$V/m.
  The initial conditions are $w_0=-0.5-i0.3$
  and $\widetilde{w}_0=0.3-i0.1$. Also shown is $b(\tau)$, corresponding to
  $w_0=0$. Note that each trajectory is a circular arc and that they all end at
  $w_\infty=1.$}\label{wtaucase2}
  \end{figure}

\noindent {\bf Case 2}\, Next consider the case $\Delta=
((cB)^2-{E}^2)/{E}^2=0\Longleftrightarrow
{E}=c{B}\;\mbox{and}\;\widetilde{B}=1$. The denominator in the
integrand of (\ref{intEB2}) is $w^2-2w+1=(w-1)^2$ and its solution
is
\begin{equation}\label{soleveqncase2}
w(\tau)=1-\frac{1}{i\Omega\tau +C}
\end{equation}
with $ C=-\frac{1}{w_0-1}.$ This s-velocity is graphed in Figure
\ref{wtaucase2}.

\begin{figure}[h!]
  \centering
 \scalebox{0.35}{\includegraphics{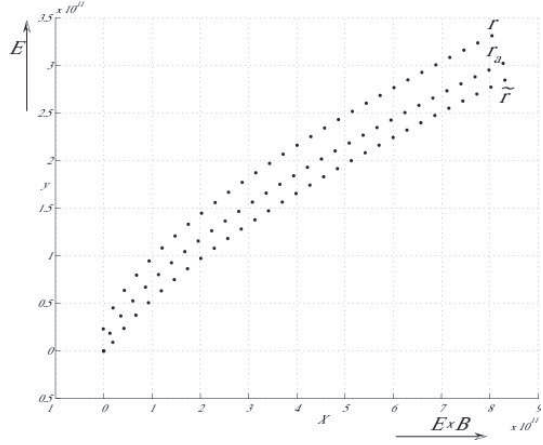}}
  \caption[Charged particle space trajectories for $|{\bf  E}|=c|{\bf  B}|$.]
  {The space trajectories $\mathbf{r}(t)$ of the test particles of Figure \ref{wtaucase2}
   during 3000 seconds.  The position of each particle is shown at fixed time intervals
  $dt=100s$.}\label{spacetrcase2}
\end{figure}

If the initial velocity is zero, $C=1$ and using that
$\gamma\mathbf{v}=\frac{2c\mathbf{w}}{1-|\mathbf{w}|^2},$ the
position of the charge as a function of the proper time is
\begin{equation}\label{proppos2}
\mathbf{r}(\tau)=2c\left(\frac{\Omega^2\tau^3}{3},\frac{\Omega\tau^2}{2}\right).
\end{equation} and the lab time as a function of the proper time
is
\begin{equation}\label{timescase2}
t(\tau)=\int_0^\tau \gamma (\tau ') d\tau '=\tau+\frac{2\Omega
^2}{3}\tau ^3.
\end{equation}
Equations  (\ref{proppos2})  and (\ref{timescase2})  give the
complete solution for this case. The space trajectories
$\mathbf{r}(t)$ of the test particles is given on of Figure
\ref{spacetrcase2}

\medskip

\noindent {\bf Case 3}\, Consider the case $\Delta= ((cB)^2-
E^2)/{E}^2<0\;\Longleftrightarrow
{E}>cB\;\mbox{or}\;\widetilde{B}<1$.

\begin{figure}[h!]
  \centering
 \scalebox{0.35}{\includegraphics{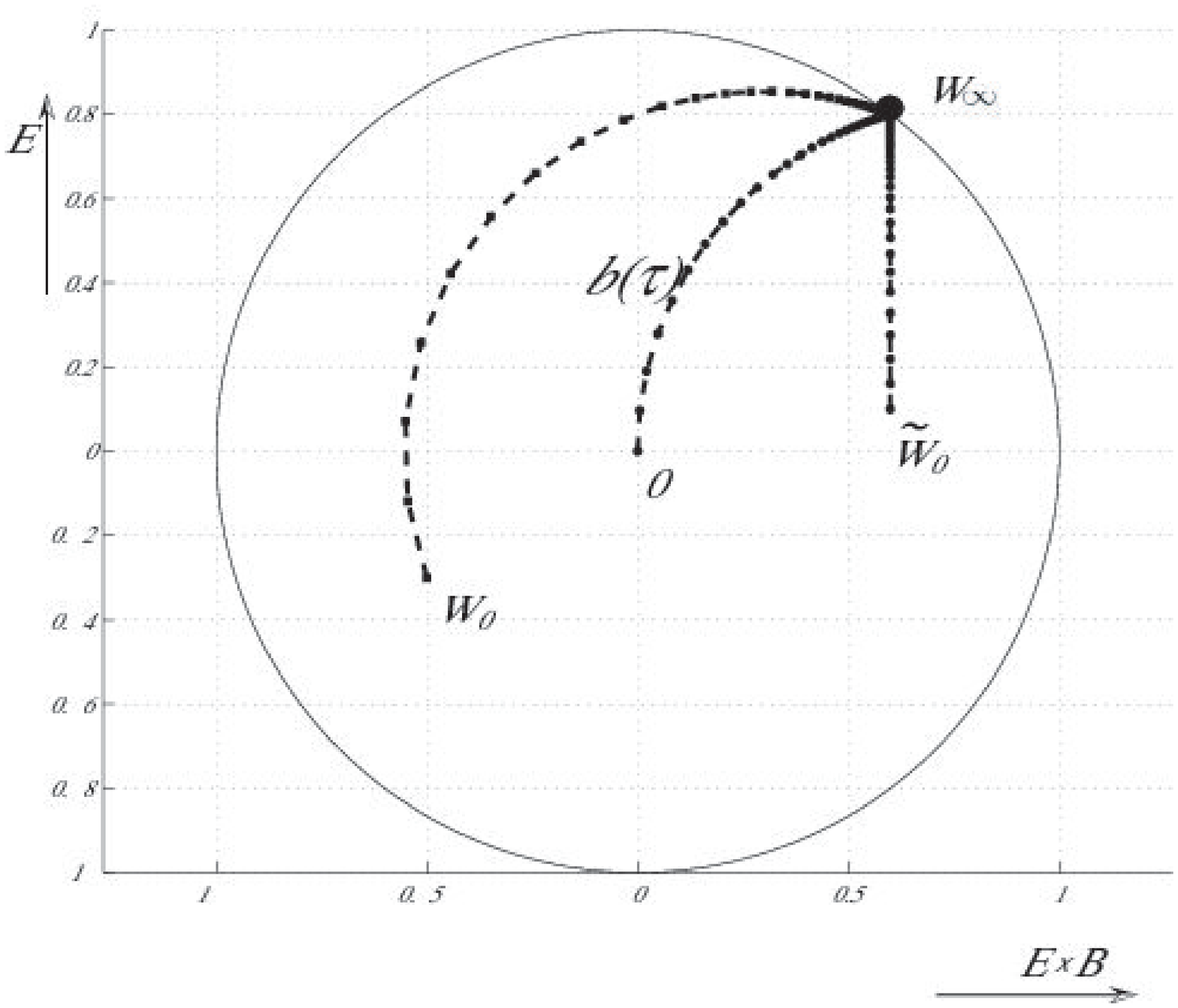}}
  \caption[Charged particle trajectories of s-velocity $w(\tau)$ for $|{\bf  E}|>c|{\bf  B}|$.]
  {The trajectories of the s-velocity $w(\tau)$ of a charged
  particle with $q/m=10^7$C/kg
  in constant, uniform fields ${\bf  E}=1$V/m and $cB=0.6$V/m.
  The initial conditions are $w_0=-0.5-i0.3$
  and $\widetilde{w}_0=0.6+i0.1$. Also shown is $b(\tau)$, corresponding
  to $w_0=0$. Note that the trajectories all end at
  $w_\infty=0.6+i0.8$.}\label{wtaucase1}
\end{figure}

Just as in Case 1, we rewrite the denominator of the integrand in
Eq. (\ref{intEB2}) as $w^2-2\widetilde{B}w+1=(w-\alpha
_1)(w-\alpha _2),$ where \begin{equation}\alpha
_1=\widetilde{B}-i\delta\;\;\mbox{and}\;\; \alpha
_2=\widetilde{B}+i\delta=\overline{\alpha}_1\end{equation}   and
$\delta=\sqrt{1-\widetilde{B}^2}>0.$ By introducing $\nu$ as in
(\ref{nu}) and an s-velocity ${w}_d
\equiv{\widetilde{B}}/\left({1+\delta}\right)$ we can write the
solution as:
\begin{equation}\label{solevoleqncase1}
w(\tau)=w _d\oplus _s (i\tanh(\nu\tau)\oplus _s \widetilde{w}_0).
\end{equation}
This s-velocity is graphed in Figure \ref{wtaucase1}

\begin{figure}[h!]
  \centering
 \scalebox{0.3}{\includegraphics{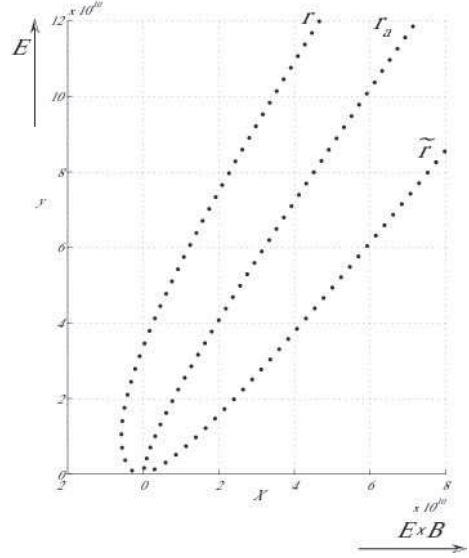}}
  \caption[Charged particle space trajectories for $|{\bf  E}|>c|{\bf  B}|$.]
  {The space trajectories $\mathbf{r}(t)$ of the test particles of Figure \ref{wtaucase1}
   in the same electromagnetic field during 500 seconds.
   The position of each particle is shown at fixed time intervals
  $dt=10s$. }\label{spacetrcase1}
\end{figure}

For the velocity of the charge we get
\begin{equation}\label{solevoleqncase1a}
\mathbf{v}(\tau)=\mathbf{v}_d\oplus _E
(c\tanh(2\nu\tau)\mathbf{j}\oplus  \widetilde{\mathbf{v}}_0 ),
\end{equation}
where $\mathbf{v}_d =(c^2 B/E)\mathbf{i}$ is the drift velocity
and $\widetilde{\mathbf{v}}_0$ is the initial velocity in the
drift frame. From this it follows that
\begin{equation}\label{finalr32}
\mathbf{r}(\tau)=\int_0^{\tau}\gamma\mathbf{v}(\tau')\,d\tau' =
\frac{\gamma_d}{\nu'}\left(\gamma_d v_d
(\sinh(\nu'\tau)-\nu'\tau),c(\cosh(\nu'\tau)-1)\right)\end{equation}
and the lab time $t$ as a function of the proper time is
\begin{equation}\label{finalt3}
 t(\tau)=\int_0^{\tau}\gamma(\tau')d \tau'=\gamma_d^2\left(\frac{\sinh(\nu'\tau)}{\nu'}-\frac{v_d^2}{c^2}\tau\right).
\end{equation}
Equations (\ref{finalr32}) and (\ref{finalt3}) together give the
complete solution for this case. The space trajectory is given in
Figure \ref{spacetrcase1}.

\section{Relativistic Dynamics of the four-velocity}

In this section we will use four-velocity instead of velocity to
describe the relativistic evolution.

To define the four velocity, we consider the Lorentz space-time
transformation between two inertial systems $K'$ and $K$ with axes
chosen to be parallel. We assume that $K'$ moves with respect to
$K$ with relative velocity $\mathbf{b}$. In order for all
the coordinates to have the same units, we  describe
an event in $K'$ by $\begin{pmatrix}  ct' \\
\mathbf{r}'\end{pmatrix}$ and by $\begin{pmatrix}  ct \\
\mathbf{r}\end{pmatrix}$ in $K$. The Lorentz transformation can be
now written as in formula (19) of Part 1 as
\begin{equation}\label{lorentzK'K}
 \left( \begin{array}{c}  ct\\ \mathbf{r}
          \end{array} \right)=
        L_\mathbf{b} \left( \begin{array}{c}  ct'\\
        \mathbf{r}'  \end{array} \right)=
             \gamma \left(
         \begin{array}{cc}
              1 & \frac{\mathbf{b}^T}{c} \\
              \frac{\mathbf{b}}{c}&  P_{\mathbf{b}}+\gamma ^{-1}(I-P_{\mathbf{b}})
          \end{array} \right)
 \left( \begin{array}{c}  ct'\\ \mathbf{r}'
          \end{array} \right),
\end{equation}
with $\gamma=\gamma(\mathbf{b})=1/\sqrt{1-|\mathbf{b}|^2/c^2}$.

Consider now the space-time evolution of the origin $O'$ of system
$K'$. This origin has space coordinate $\mathbf{r}'=0$ and thus by
(\ref{lorentzK'K}) its evolution in $K$ is given by
\begin{equation}\label{four velocitytrace}
  \left( \begin{array}{c}  ct\\ \mathbf{r}
          \end{array} \right)=\gamma (\mathbf{b})\left(
         \begin{array}{c}   c  \\ \mathbf{b} \end{array}
          \right)t'.
\end{equation}
This shows that $O'$ moves with uniform proper velocity $\gamma
(\mathbf{b})\left(  \begin{array}{c}   c  \\ \mathbf{b}
\end{array}          \right)$ in $K$, which is called the
\textit{four-velocity} corresponding to $\mathbf{b}$, which we
will denote by $\tilde{b}$. In other words
\begin{equation}\label{four velocity notation}
  \tilde{b}=\gamma (\mathbf{b})\left(
         \begin{array}{c}   c  \\ \mathbf{b} \end{array}
          \right),
\end{equation}
which is a four-dimensional vector. The four-velocity expresses
not only the change of the position of an object but also the
change of the time rate of the clock comoving with the object.

Here too, we will assume the $EP^2$ principle, which implies that
the acceleration of an object under a given force in an inertial
system $K$ is equivalent to free motion in system $K'$ moving with
a variable relative velocity $\mathbf{b}(t)$ with respect to $K$.
Also here we may assume that  $\mathbf{b}(0)=0.$ Denote the
initial velocity of the object in $K'$ by $\mathbf{v}$. Since the
motion of the object in system $K'$ is free, the velocity of the
object will remain constant in $K'$. The four-velocity
$\tilde{v}=\gamma (\mathbf{v})\left(
\begin{array}{c}   c  \\ \mathbf{v} \end{array} \right)$
will also remain constant. We denote the proper time of the object
by $\tau$. By use of (\ref{lorentzK'K}) and (\ref{four
velocitytrace}) and the well-known formulas (see \cite{F04}) for
relativistic velocity addition and the transformation of the
corresponding $\gamma$'s, we can calculate the world-line of the
the object in $K$ as
\[ \left( \begin{array}{c}  ct\\ \mathbf{r}
          \end{array} \right)=
        L_{\mathbf{b}} \left( \begin{array}{c}  ct'\\
        \mathbf{r}'  \end{array} \right)=L_{\mathbf{b}(t)}\tilde{v}\tau=
        L_{\mathbf{b}}\gamma (\mathbf{v})\left(
\begin{array}{c}   c  \\ \mathbf{v} \end{array} \right)\tau=\]
    \[ =  \gamma(\mathbf{b})
  \gamma (\mathbf{v}) \left(
         \begin{array}{cc}
              1 & \frac{\mathbf{b}^T}{c} \\
              \frac{\mathbf{b}}{c}&  P_{\mathbf{b}}+\gamma(\mathbf{b})^{-1}(I-P_{\mathbf{b}})
          \end{array} \right)
 \left( \begin{array}{c}  c\\ \mathbf{v}
          \end{array} \right)\tau=\]\[
  =\gamma(\mathbf{b})
  \gamma (\mathbf{v}) \left(
         \begin{array}{c}
              c+ \frac{\mathbf{b}\cdot \mathbf{v}}{c} \\
              \mathbf{b}+  P_{\mathbf{b}}+\gamma(\mathbf{b})^{-1}(I-P_{\mathbf{b}})
          \end{array} \right)\tau=\]
\begin{equation}\label{four vel addition ini}
=\gamma(\mathbf{b})
  \gamma (\mathbf{v})(1+ \frac{\mathbf{b}\cdot \mathbf{v}}{c^2})
 \left( \begin{array}{c}
              c \\
              \mathbf{b}\oplus \mathbf{v}\end{array} \right)\tau
              =\gamma(\mathbf{b}\oplus \mathbf{v})
 \left( \begin{array}{c}
              c \\
              \mathbf{b}\oplus \mathbf{v}\end{array} \right)\tau.\end{equation}
This shows that
$L_{\mathbf{b}(t)}\tilde{v}=\widetilde{\mathbf{b}\oplus
\mathbf{v}}$ and that the four velocity transformation between
$K'$ and $K$ is given by multiplication by the $4\times 4$ matrix
of $ L_{\mathbf{b}}$.

As a result, the relativistic acceleration, which is the generator
of the four-velocity changes, is obtained by differentiating the
matrix of $ L_{\mathbf{b}(t)}$ with respect to $t$ at $t=0$. Since
$\mathbf{b}(0)=0$, we have  $\gamma(\mathbf{b}(0))=1$ and $
\frac{d}{dt}\gamma(\mathbf{b}(t))\Bigr|_{t=0}=0$, Denoting $
\frac{d}{dt}\mathbf{b}(t)\Bigr|_{t=0}=\mathbf{a},$ we get the
matrix for \textit{relativistic acceleration}
\begin{equation}\label{relat accel def}
  \delta
  (L_{\mathbf{b}(t)})=\frac{d}{dt}L_{\mathbf{b}(t)}\Bigr|_{t=0}=\frac{1}{c}\left(
         \begin{array}{cc}
              0 & \mathbf{a}^T \\
              \mathbf{a}  &0
          \end{array} \right).
\end{equation}
Using the fact that for small velocities
$\mathbf{f}=m_0\mathbf{a}$, by Newton's dynamic law, the
four-velocity and relativistic acceleration in special relativity
 become:
\begin{equation}\label{relat fouvel mechanic}
  m_0\frac{1}{c}\left(
         \begin{array}{cc}
              0 & \mathbf{a}^T \\
              \mathbf{a}  &0
          \end{array} \right)\gamma(\mathbf{v})
 \left( \begin{array}{c}
              c \\
               \mathbf{v}\end{array} \right)=m_0\gamma(\mathbf{v})
 \left( \begin{array}{c}
              \frac{\mathbf{a}\cdot\mathbf{v}}{c} \\
               \mathbf{a}\end{array} \right)=\gamma(\mathbf{v})
 \left( \begin{array}{c}
              \frac{\mathbf{f}\cdot\mathbf{v}}{c} \\
               \mathbf{f}\end{array} \right).
\end{equation}
The last expression is called the \textit{four-force}, see
\cite{rindler} p. 123.

General four-velocity transformations between two inertial systems
also include rotations, which can be expressed by a $3\times 3$
orthogonal matrices $U(t)$. We will extend such a matrix to a
$4\times 4$ matrix by adding zeros in the time components outside
the diagonal and assume that $T(0)=I$. The general four-velocity
transformation will then be
\begin{equation}\label{genera fourvel trans}
  T(t)=L_{\mathbf{b}(t)}U(t)
\end{equation}
and its generator, representing \textit{relativistic
acceleration}, is
\begin{equation}\label{generator gen fourvel}
  \delta
  (T(t))=\frac{d}{dt}T(t)\Bigr|_{t=0}=
  \frac{d}{dt}L_{\mathbf{b}(t)}\Bigr|_{t=0}+\frac{d}{dt}U(t)\Bigr|_{t=0}=\frac{1}{c}\left(
         \begin{array}{cc}
              0 & \mathbf{a}^T \\
              \mathbf{a}  &A \end{array} \right),
\end{equation}
where $A=\frac{d}{dt}T(t)\Bigr|_{t=0}$ is a $3\times 3$
antisymmetric matrix.

We have seen that relativistic acceleration includes both linear
and rotational acceleration and is a linear map on the
four-velocities. The matrix representing the relativistic
acceleration is antisymmetric if both indices are space indices or
both are time indices and is symmetric if one of the indices is
spacial and the other is a temporal. Moreover, any
\textit{relativistic force}, which is a multiple of the
relativistic acceleration by $m_0$, must have the
form of the
electromagnetic tensor
\begin{equation}\label{electromag tensor}
  \hat{F}=\frac{q}{c}\left(
         \begin{array}{cccc}
              0 & E_1&E_2&E_3\\E_1&0&B_3&-B_2\\
              E_2&-B_3&0&B_1\\E_3&B_2&-B_1&0
               \end{array} \right)
\end{equation}
and must transform from one inertial system to another in the same
way that this tensor transforms. The electromagnetic dynamic
equation in our notation is
\begin{equation}\label{electrodynam eq}
  m_0\frac{d}{d\tau}\tilde{v}=\hat{F}\tilde{v}.
\end{equation}

 In classical mechanics, a
force was represented by a differential one-form which expressed
the change of the velocity and space displacement of the object in
the direction of the force. In special relativity, a force (a
non-rotating one) causes more change then it causes in classical
mechanics. It  also causes a change in the rate of a clock
connected to the object due to the change of the magnitude of the
object's velocity. Thus, it has to be represented by a
differential two-form. On the other hand, forces causing rotation,
like the magnetic force need to be described by differential
two-forms also in classical mechanics. Thus, only in relativistic
dynamics can these two forces be combined effectively as a single
force.

\section{Discussion}

We have shown that an analog of the Equivalence Principle leads to
the known relativistic dynamic equation. The relativistic force is
defined by an element of the Lie algebra of the group $Aut
_p(D_v)$ of projective automorphisms of the ball of
relativistically admissible velocities $D_v$. This Lie algebra is
a quadratic polynomial on $D_v$ where the constant and quadratic
coefficients define an analog of electric force, while the linear
term corresponds to a magnetic force. Such decomposition exists for
any force in relativity. The Lie algebra $aut _p(D_v)$ is
described by the triple product associated with the domain $D_v$
which is a domain of type I in Cartan's classification.

The relativistic force on a  new dynamic variable - symmetric
velocity- is an element of $aut _c(D_s)$- the Lie algebra of the
conformal group on the  ball of relativistically admissible
symmetric velocities $D_s$. For velocities with the speed of light,
the symmetric velocity and the regular velocity are equal. This
explains the known fact that the Maxwell equations (related to
electro-magnetic propagation with the speed of light) are invariant
under the conformal group. But in order to obtain conformal
transformation for massive particles we must use symmetric
velocity instead of the regular velocity. The use of symmetric
velocity helps to find analytic solutions for relativistic
dynamic equations.

The Lie algebra $aut _c(D_s)$ is described by the triple product
associated with the domain $D_s$.  In this case, this  is a
domain of type IV in Cartan's classification, called the Spin
factor. A complexification of this domain leads to Dirac bispinors,
an analog of the geometric product of Clifford algebras. We also
obtain both spin 1 and spin $1/2$ representations of the Lorentz
group on this domain, see \cite{F05}. This may provide a
connection between Relativity and Quantum Mechanics.

By applying the analog of the Equivalence Principle to the
four-velocity we showed that the relativistic dynamics equation
leads to an analog of the electro-magnetic tensor.

We want to thank Dr. Tzvi Scaar and Michael Danziger for helpful
remarks.

\end{document}